\documentstyle[12pt]{article}
\renewcommand{\baselinestretch}{1.4}
\newcommand{\be}{\begin{equation}}
\newcommand{\ee}{\end{equation}}
\newcommand{\ba}{\begin{array}{l}}
\newcommand{\ea}{\end{array}}
\newcommand{\banonum}{\begin{eqnarray*}}
\newcommand{\eanonum}{\end{eqnarray*}}
\newcommand{\baa}{\begin{eqnarray}}
\newcommand{\eaa}{\end{eqnarray}}

\date{}
\title{INELASTIC PROCESSES IN THE COLLISION OF RELATIVISTIC
HIGHLY CHARGED IONS WITH ATOMS}
\author{V.I.MATVEEV \thanks{e-mail: matveev@vict.silk.org} ,
 Kh.Yu.RAKHIMOV\\ AND\\ D.U.MATRASULOV \\
Heat Physics Department of the Uzbek Academy of Sciences, \\
28 Katartal St., 700135 Tashkent, Uzbekistan }

\begin{document}
\large
\maketitle
\begin{abstract}

A general expression  for the cross
sections of inelastic collisions of fast (including relativistic)
multicharged ions with atoms which is based on the genelazition of
the eikonal approximation is derived.
This expression is applicable
for wide range of
collision energy and has the standard nonrelativistic
limit and in the ultrarelativistic limit coincides with the Baltz's exact
solution ~\cite{art13} of the Dirac equation. As an application of the obtained result
the following processes are calculated:
the excitation and ionization cross sections of hydrogenlike atom;
the single and double excitation and ionization of heliumlike atom;
the multiply ionization  of neon  and argon atoms;
the probability and cross section of
K-vacancy production in the relativistic $U^{92+} - U^{91+}$ collision.
The simple analytic formulae for the cross
sections of inelastic collisions and the recurrence
relations between the ionization cross sections of different
multiplicities are also obtained. Comparison of our results with
the experimental data and the results of other calculations are given.
\end{abstract}

\section*{Introduction}
Study of inelastic processes in the collision of atoms with fast (including
 relativistic)  highly charged ions has been a subject of considerable interest
 recently. One of motivation for this interest is the anticipated large
effective field strength creating  by highly charged ions which exceeds internal
atomic field strength. Creating such a fields by the another methods is
very complicated problem presently. Thus up to now, collision experiments
with highly charged ions are the only way for investigating atoms and molecules
in superstrong fields. From the fundamental viewpoint study of behavior of
matter in superstrong electromagnetic fields is one of the important problems
of modern physics. Furthermore, a number of applied problems such as measuring
of energy spectrum of nuclear fission, synthesis of superheavy elements,
interpretation of data on superheavy cosmic rays, ion diagnostics and spectroscopy
of plasma, etc., stimulate  the extensive study of collisions of highly charged ions
with atoms. Among such a processes collisions of fast and relativistic highly
charged ions with atoms are of great interest due to the anticipated large
cross sections. Therefore these processes are also of applied importance.

The theoretical methods used for relativistic collisions are very similar
to those applied for nonrelativistic collision energies and have mainly
the perturbative character. The relativistic treatment within the plan wave
Born approximation dates back to M\" oller ~\cite{a1} and has been further
developed by Jamnikand Zupancic ~\cite{a2} , Davidovich et.al
~\cite{a3} and by Anholt ~\cite{a4}.
As is well known ~\cite{art4}  in a relativistic ion -atom collisions
perturbation theory begins to break down
for large projectile charges (for $Z\geq 75$).
For example the well known Born approximation leads to a result in which
(for small impact parameters) the ionization probability exceeds unity.
For this reason nonperturbative methods for calculation
of such processes are needed. Presently a few nonperturbative results are available.
Becker et.al. ~\cite{art12}  used a finite difference method to solve the Dirac
equation for $U^{92+} - U^{91+}$ collisions at 1 GeV/u on a discretized grid.
Recently Baltz ~\cite{art13}  have obtained exact solution of the Dirac equation
for relativistic heavy ion collisions in the ultrarelativistic limit.
Another nonperturbative methods
of calculation of cross section such inelastic
processes are the Glauber approximation ~\cite{art9,art3,khab98} and
sudden approximation ~\cite{art5,art6,a5}.
In this work using the
eikonal approximation we obtain a general expression for the cross section
of inelastic collision of fast as well as relativistic highly charged ion
with complex atom.
This expression is applicable
for wide range of
collision energies and has the standard nonrelativistic
limit and in the ultrarelativistic limit coincides with the Baltz's exact
solution ~\cite{art13} of the time-dependent Dirac equation. As an application
of the obtained result excitation and ionization  cross sections of hydrogenlike atom, single
and double excitation and ionization of heliumlike atom, multiply
ionization (up to eightfold) of neon atom and (up to eighteen) of
argon atom, probability and cross section of
the K-vacancy production in the relativistic $U^{92+} - U^{91+}$ collision are
calculated.
The simple analytic formulae for the cross
sections of inelastic collisions and the recurrence
relations between the ionization cross sections of different
multiplicities are also obtained. Comparison of our results with
the experimental data as well as the results of other calculations are given.

\section*{Eikonal approximation for relativistic ion atom collisions}

A general expression for the inelastic transition amplitude  from state
$\mid\Phi_{i}>$  to state $\mid\Phi_{f}>$ for the collision of
relativistic highly charged ion with light (nonrelativistic before and after collision) atom
in the Glauber approximation has been obtained previously
~\cite{art11} (following ~\cite{art14} ):
\begin{eqnarray}
\displaystyle f_{if}({\bf q})=\frac{ik_{i}}{2\pi}\int e^{-i{\bf qb}}
<\Phi_{f}\mid\left[1-exp\{-\frac{i}{v}\int Udx\}\right]\mid \Phi_{i}>d^{2}b\,,
\label{eq:af1}
\end{eqnarray}
where ${\bf q}={\bf k}_{f}-{\bf k}_{i}$ is the momentum transfer.
The scattering potential $U=U(x,{\bf b};\{{\bf r}_{a}\})$ is a function of
ion's coordinates ${\bf R}=(x,{\bf b})$ as well as coordinates of
atomic electrons, which we denote as $\{{\bf r}_{a}\}$,
$\;\;a=1,2,...N$, where $N$ is the number of electrons.

To generalize this eikonal approximation for the case of relativistic ion-heavy
(relativistic) atom collision one should account the followings:\\
a) behavior of atomic electrons are described by the Dirac equation;\\
b) in the Glauber approximation  $U(x,{\bf b};\{{\bf r'}_{a}\})$ is the
static Coulomb potential which is induced by the atomic nuclear and the
electrons which are in fixed (and simultaneous from the projectile viewpoint)
points ${\bf r}'_{a} = (x'_{a},y'_{a},z'_{a})$. Then we have
\begin{eqnarray}
\displaystyle
\frac{1}{v}\int \limits^{+\infty}_{-\infty}Udx
=\sum ^{N}_{a=1} \chi_{a}({\bf b},{\bf s'}_{a})\;,\;\;
\chi_{a}({\bf b},{\bf s'}_{a})
=\frac{2Z}{v}ln\frac{\mid{\bf b}-{\bf s'}_{a}\mid}{b},
\nonumber\end{eqnarray}
axis $x$ is directed on ${\bf k}_{i}$, ${\bf s'}_{a}=(y'_{a},z'_{a})$
is the two-dimensional vector. Let us, for the definiteness, consider
electrons in the instantaneous positions  ${\bf r'}_{a}$ at the moment $t'=0$ in the rest frame
of the ion and corresponding wave function is  $\Psi'({\bf r'_{a}},t')$.
Then from (~\ref{eq:af1}) we have

\begin{eqnarray}
\displaystyle
f_{if}({\bf q})=\frac{ik_{i}}{2\pi}
\int \Psi'^{+}_{f}({\bf r'_{a}},t'=0) \left[1-exp\{-\frac{i}{v}
\int U(x,{\bf b};\{{\bf r'}_{a}\})dx\}\right] \times\nonumber\\
\displaystyle\times \Psi'_{i}({\bf r'_{a}},t'=0)
exp(-i{\bf q}{\bf b})d^{2}b\prod^{N}_{a=1}d^{3}r'_{a}\;\;.
\nonumber\end{eqnarray}

In the rest frame of the atom we have for $t'=0$:
$$
x_{a} = \gamma x'_{a}\;,\;\;
{\bf s}_{a}={\bf s'}_{a}\;,\;\;
t=-x_{a} \frac{v}{c^2};
$$
$$
\Psi({\bf r_{a}},t)= \psi({\bf r_{a}})exp(-iEt)=
\psi({\bf r_{a}})exp(iEx_{a} \frac{v}{c^2})
= S^{-1}_{a}\Psi'({\bf r'_{a}},t'=0);
$$
$$
d^{3}r_{a}= dx_{a}dy_{a}dz_{a}=
\gamma d^{3}r'_{a}=
\gamma dx'_{a}dy'_{a}dz'_{a}\;,
$$
where $\gamma=1/\sqrt {1-v^2/c^2}$,  $S_{a}^{-1}$  is the Lorentz matrix
which transforms the wave function from ion-to-atom rest frame. It acts only
to  byspinor indices of the atomic electron with index $a$ (corresponding
Dirac matrices are ${\bf {\alpha}}_{a}$),  $S_{a}^{-2}$ is the  matrix
which is written as ~\cite{art1}
$$
S^{-2}_{a}=\gamma
(1-{\bf v}{\bf {\alpha}}_{a}/c)\;.
$$
Thus in the rest frame of the atom the transition amplitude from state
 $\mid\psi_{i}>$ with energy $E_{i}$ to state  $\mid\psi_{f}>$ with energy
$E_{f}$ can be written, in the Glauber
approximation, as the following:
\begin{eqnarray}
\displaystyle
f_{if}({\bf q})=
\frac{ik_{i}}{2\pi}
\int
<\psi_{f}\mid[1-exp\{-i\sum ^{N}_{a=1} \chi_{a}({\bf b},{\bf s}_{a})\}]
\times\nonumber\\ \displaystyle\times
\gamma^{-N}S^{-2} exp{[i\sum_{a} \frac{vx_{a}}{c^2}(E_{f}-E_{i})]}\mid\psi_{i}>
exp(-i{\bf q}{\bf b})d^{2}b,
\label{eq:af2}
\end{eqnarray}
where $S^{-2}=\prod^{N}_{a=1} S^{-2}_{a}$. This is the final expression
for the transition amplitude which could be used in the collision of
relativistic ion with complex (relativistic or nonrelativistic) atom.
The formula (~\ref{eq:af2}) (as well as (~\ref{eq:af1}))
is applicable when the collision time is considerably less than characteristic atomic time.

If we are not interesting in ion's scattering  angles
one can perform integrating over these angles. So, for small angles one has
\be
d\Omega\approx d^2q/(k_{i}k_{f})\approx d^{2}q/k^{2}.
\label{angl}
\ee
Representing $\mid f_{if}({\bf q})\mid^2$ from (~\ref{eq:af2}) in the form of product
of integrals over  $d^{2}b'$ and $d^{2}b$ and performing integration over these
variables  by using eq.(~\ref{angl})
and integral representation of the $\delta$-function we obtain the cross
section of transition from state $\mid\psi_{i}>$ to state $\mid\psi_{f}>$
for the relativistic ion-atom collision:
\begin{eqnarray}
\displaystyle
\sigma=\int
d^{2}b\mid
<\psi_{f}\mid[1-exp\{-i\sum ^{N}_{a=1} \chi_{a}({\bf b},{\bf s}_{a})\}]
\times\nonumber\\ \displaystyle\times
\gamma^{-N}S^{-2} exp{[i\sum_{a} \frac{vx_{a}}{c^2}(E_{f}-E_{i})]}\mid\psi_{i}>
\mid^{2}\:.
\label{eq:af3}
\end{eqnarray}
In this expression integrand is interpreted as the
transition probability
from state $\mid\psi_{i}>$ to state $\mid\psi_{f}>$ in the collision with the
impact parameter ${\bf b}$. One should note that in this form this probability
coincides with exact one obtained by Baltz ~\cite{art13}  for the
ultrarelativistic case and has a standard nonrelativistic limit ~\cite{art14}.
For long-range potentials integral in (\ref{eq:af3})
diverges for large impact parameters. However, as is known, such a divergence
is not considerable ~\cite{art3,art7}, since for large impact parameters the
Born approximation can be applied. There is a region in which the regions
of applicability of Born and eikonal approximations overlap one which other.
This allows a correct matching of cross sections over the impact parameter.

Consider this matching in the case of K-vacancy production in the collision of
relativistic heavy ion with heavy atoms when the transition of the atomic
K-shell-electron from the state $|i>$  to the continuum state  $|{\bf k}>$ with
momentum  ${\bf k}$ will occurs. Let's denote via $b_{0}$ upper integration
limit over the impact parameter $b$ in (\ref{eq:af3}). For ${b\gg s}$
 and orthogonal  $|{\bf k}>$ and $ |i> $
the generalized inelastic formfactor

\be
<f|1-exp\{-i\frac{2Z}{v} ln\frac{\mid{\bf b}-{\bf s}\mid}{b}\}|i>\approx
<f|exp\{i{\bf q}{\bf r}\}|i>\;
\label{eq:af4}
\ee
tends  (for small ${\bf q}$) to $i {\bf q} <f|{\bf r}|i>$,
where ${\bf q}=2Z{\bf b}/(vb^2)$.
Therefore integral in (~\ref{eq:af3}) over $d^{2}b$ depends on $b_{0}$
logarithmically and for this reason the contribution of the region $b<b_{0}$
to the cross section can be written as

\be
\sigma(b<b_0)=8\pi\frac{Z^2}{v^2}\lambda_{i}
 \ln\frac{2\alpha_{i}}{q_0},\;\; q_{0}=2Z/(v b_{0}),
\label{eq:af5}
\ee
where
\be
\lambda_{i}
=\int d^3k|<{\bf k}|{\bf r}|i>|^2/3,
\label{lambda}
\ee
\begin{eqnarray}
\displaystyle
\alpha_{i}=\lim_{b_{0}\rightarrow \infty}\:\frac{Z}{vb_0}exp
\biggl\{\frac{1}{\lambda_{i}}\frac{v^2}{8\pi Z^2}\int_{0}^{b_{0}}
2\pi bdb\int d^3k
\times\nonumber\\ \nonumber\\ \displaystyle\times
\mid<{\bf k}|[1-exp(-i\frac{2Z}{v} ln\frac{\mid{\bf b}-{\bf s}\mid}{b})]
\gamma^{-N}S^{-2} exp[i\frac{vx}{c^2}(E_{k}-E_{i})]|i>\mid^2 \biggl\}.
\label{alpha}
\end{eqnarray}

In the region $b>b_0$  the field of the ion is a weak perturbation and one
can use the so-called Bethe asymptotic:
\be
\sigma_{i}(b>b_{0})=8\pi\frac{Z^{2}}{v^{2}} \lambda_{i}
\left(
\ln\frac{2v}{\eta b_{0}\omega_{i}\sqrt{1-\beta^{2}}} -
\frac{\beta^{2}}{2}
\right)\;.
\label{eq:af6}
\ee
where  $\eta=e^{B}=1.781$ \,\,(B=0.5772 is the Euler constant),
$\omega_{i}$ is the "average" ionization frequency :
\be
\ln \omega_i = \frac{\int d^3 k |<{\bf k}|{\bf r}|i>|^2 ln \Omega_{{\bf
k}i}}{\int d^3 k |<{\bf k}|{\bf r}|i>|^2},
\label{eq:af7}
\ee
where $\Omega_{{\bf k}i}=\epsilon_{\bf k}-\epsilon_i$ - transition frequency.
Summing (~\ref{eq:af5}) and (~\ref{eq:af6}) we obtain the total K-shell ionization
cross section:
\be
\sigma_{i}=8\pi\frac{Z^{2}}{v^{2}} \lambda_{i}
\left(
\ln\frac{2\alpha_{i} v^{2}}{\eta Z\omega_{i}\sqrt{1-\beta^{2}}} -
\frac{\beta^{2}}{2} \right)\;.
\label{eq:afsig}
\ee
Quantities $\lambda_{i}$,  $\alpha_{i}$ and $\omega_{i}$ are calculated
numerically using the formulae (~\ref{lambda}), (~\ref{alpha}) and (~\ref{eq:af7}).
Note that the dependence on the cut-off parameter $b_{0}$ disappears
after matching.

If the states of more than one electron
change after the collision, or the dipole transitions are forbidden, integration
over the impact parameter in (~\ref{eq:af3}) can be extended to the hole
plan of impact parameters (since integrand garantees convergence) and
there is no need in matching.

Obtained formulae are of the general character and can be applied to the
collisions of atoms with ions of arbitrary charges. Specificity of the
collisions
of highly charged ions with atoms is the fact that the cross sections of such
collisions are large enough (considerably exceeding atomic sizes). This enables
one to use for the calculations of the cross section the large impact parameter
approximation (~\ref{eq:af4}). In this case the procedure of mathcing is most simply applicable  for
the collsions of relativistic ions
with nonrelativistic (before and after collision
\footnote{Strictly speaking, atomic electrons appearing in the
continuum in the result of ionization by the impact of relativistic
ion can take relativistic velocities. However, as is shown in ~\cite{art3}
such a processes occur for the small impact parameters and corresponding
contributions to the full ionization cross section by the impact of
highly  charged ion can be neglected.
})
atoms. In this case
in  (~\ref{eq:af3}) $\psi_{i}$ и $\psi_{f}$  are the two-component
spinors and  $\gamma^{-N}S^{-2}=1$. Besides that
$exp[i\sum_{a} x_{a}v(E_{f}-E_{i})/c^2]=1,$
\begin{eqnarray}
\alpha_{i}=\lim_{q_{0}\rightarrow 0}\:\frac{q_0}{2}exp
\left\{\frac{1}{\lambda_{i}}\int_{q_0}^{\infty}\frac{dq}{q^3}\int d^3k
|<{\bf k}|exp(-i{\bf q\,\bf r}\,)|i>|^2 \right\},
q_0=\frac{2Z}{vb_0}
\label{alphan}
\end{eqnarray}
and $\omega_{i}$
depend on only the atomic characteristics but not depend on the
impact parameter, projectile charge and velocity:
\begin{eqnarray}
\omega_{i}(Z_a)&=&\omega_{i}(Z_a=1)Z_a^2,\;\;
\nonumber\\
\alpha_{i}(Z_a)&=&\alpha_{i}(Z_a=1),\;\;
\nonumber\\
\lambda_{i}(Z_a)&=&\lambda_{i}(Z_a=1)/Z_a^2\;.
\label{n}
\end{eqnarray}

These facts allow us to obtain simple analitical expressions for the cross sections
by matching.
\section*{Collisions with hydrogenlike atoms}

Here we give the trasitions cross sections (obtained by matching using
the large impact parameter approximation (~\ref{eq:af4}))
of nonrelativistic hydrogenlike
atom (with the nuclear charge$Z_{a}$) from the ground state
to the state with principal quantum number $n$ in the
collision  with relativistic highly charged ion
\be
\sigma_n =\frac{2^{11}\pi}{3}\frac{Z^2}{v^2}\frac{n^7}{(n^2-1)^5}
\left(\frac{n-1}{n+1}\right)^{2n}\frac{1}{Z^2_a}
\left\{\ln\left(\frac{\gamma_n v^2 Z_a}{Z\Omega_n
\sqrt{1-\beta^{2}}}\right) -\frac{\beta^2}{2}\right\},
\label{eq:af9}
\ee
where  $\Omega_{n}=\epsilon_{n}-\epsilon_1$,
some of $\gamma_n$ equal
$$
\ba
\gamma_2=0.30;\;\; \gamma_3=0.44;\;\; \gamma_4=0.49;\;\; \gamma_5=0.53;\;\;
\gamma_6=0.54;\\
\gamma_7=0.55;\;\; \gamma_8=0.56;\;\; \gamma_9=0.57;\;\; \gamma_{10}=0.57;\;\;
\gamma_{11}=0.57.
\ea
$$
For the total ionization cross section we have
\be
\sigma_i =8\pi \frac{Z^2}{v^2} 0.283\frac{1}{Z^2_a}
\left\{\ln\left(\frac{5.08v^2}{Z Z_a\sqrt{1-\beta^{2}}}\right)
-\frac{\beta^2}{2}\right\} .
\label{eq:af10}
\ee
Summing (~\ref{eq:af9}) over all $n$ one obtains the total cross section of
excitation of discrete states
\be \sigma_{ex}=\sum_{n=2}^{\infty}\sigma_n=8\pi\frac{Z^2}{v^2} 0.717\frac{1}{Z^2_a}
\left\{\ln\left(\frac{0.84v^2}{Z Z_a\sqrt{1-\beta^{2}}}\right)
-\frac{\beta^2}{2}\right\},
\label{eq:af11}
\ee
and the total inelastic cross section
\be
\sigma_r =\sigma_{ex} +\sigma_i =8\pi \frac{Z^2}{v^2} \frac{1}{Z^2_a}
\left\{\ln\left(\frac{1.4v^2}{Z Z_a\sqrt{1-\beta^{2}}}\right)
-\frac{\beta^2}{2}\right\} .
\label{eq:af12}
\ee
Obtained formulae (~\ref{eq:af9}) and (~\ref{eq:af10}) can be used
for the estimation the cross sections of excitation and ionization
of K-shell by the impact of relativistic highly charged ion when the K-shell
electrons can be described by the hydrogenlike wave functions with the effective
charge $Z_{a}$. For the estimation of L-shell
excitation or ionization  cross sections one can use excitation or ionization
cross sections of hydrogenlike $2s$ and $2p$ states. For the
hydrogenlike atom in the initial $2s$-state we have
\be
\sigma_i =8 \pi \frac{Z^2}{v^2} 0.82\left\{\ln\left(\frac{17.1v^2}{Z
Z_a\sqrt{1-\beta^{2}}}\right) -\frac{\beta^2}{2}\right\},
\label{eq:af13}
\ee

\be
\sigma_{n}=8\pi\frac{Z^2}{v^2} \frac{2^{17}}{3} n^7
\frac{(n-2)^{2n-6}}{(n+2)^{2n+6}} (n^2-1)
\left\{\ln\left(\frac{\beta_n v^2Z_a}{Z\Omega_n \sqrt{1-\beta^{2}}}\right)
-\frac{\beta^2}{2}\right\},
\label{eq:af14}
\ee
where $n\geq 3,$ $\Omega_{n}=\epsilon_{n}-\epsilon_2$, the numbers $\beta_n$ are
$$
\ba
\beta_3=0.18;\;\; \beta_4=0.28;\;\; \beta_5=0.34;\;\; \beta_6=0.39;\\
\beta_7=0.41;\;\; \beta_8=0.42;\;\; \beta_9=0.44;\;\; \beta_{10}=0.45;\;\;
\beta_{11}=0.46.
\ea
$$

For the hydrogenlike atoms initially being in the $2p$-states
(after averaging over the projection of the angular moment of this state)
we have
\be
\sigma_i =8 \pi \frac{Z^2}{v^2} 0.53\left\{\ln\left(\frac{271v^2}{Z
Z_a\sqrt{1-\beta^{2}}}\right) -\frac{\beta^2}{2}\right\},
\label{eq:af15}
\ee

\be
\sigma_{n}=8\pi\frac{Z^2}{v^2} \frac{2^{15}}{3} n^{11}
\frac{(n-2)^{2n-7}}{(n+2)^{2n+7}} \left(\frac{11}{3}-\frac{4}{n^2}\right)
\left\{\ln\left(\frac{\beta_n v^2 Z_a}{Z\Omega_n
\sqrt{1-\beta^{2}}}\right) -\frac{\beta^2}{2}\right\},
\label{eq:af16}
\ee where $n \neq 2$,
the numbers $\beta_n$ are
$$
\ba
\beta_1=0.27;\;\; \beta_3=0.13;\;\; \beta_4=0.30;\;\; \beta_5=0.46;\;\;
\beta_6=0.58;\\
\beta_7=0.67;\;\; \beta_8=0.73;\;\; \beta_9=0.79;\;\; \beta_{10}=0.82;\;\;
\beta_{11}=0.85.
\ea
$$

To show the character of obtaining by the matching results
we give in fig.1 the ionization cross sections obtained:
in the Born approximation - 1;
by matching (formula (~\ref{eq:af10}))- 2; in the Glauber approximation
~\cite{art9} - 3; in the sudden approximation ~\cite{art7} - 4.
As is seen from this figure formula (~\ref{eq:af10}) in the region
of its
applicability ($v  \sim  Z$) is  close  enough  to  one obtained in the
Glauber approximation and tends to the Born
approximation by increasing of $v$.
Note that obtained in this section expressions for the cross section not
coincide (for $Z\sim v\ll 1$) with the results of Born approximation
$\sigma_{B}$ (this fact is the general property of the approximation (~\ref{eq:af3})),
the ratio $(\sigma_{B}-\sigma)/\sigma\rightarrow 0$ for $v \rightarrow с$.
In the region $Z\sim v\sim 1$ our results based on unitary approximation
(~\ref{eq:af3}) as the Glauber approximation give the better accordance
with the experiment in comparison with the Born approximation which, as
is  well known, is not unitary and exceeds (approximately 1.5 times)
inelastic cross section. Differing from the Glauber approximation in the
form (~\ref{eq:af2}),
~\cite{art9,art10} which
requires considerable numerical computations, our results are obtained in the
analytical form.

\section*{Excitation and ionization of nonrelativistic heliumlike atom}

When one uses the perturbation theory for the calculations of inelastic cross sections
of collisions of fast charged particles with complex atom the one-electron
excitation and ionization are the first-order effects. The
two-electron transitions are the second order effects of perturbation
theory, when the interaction of projectile with atomic
electrons and interelectron interaction are taken into account once.

Analogously other many-electron transitions can be calculated: i.e.,
the interaction of projectile with atomic electrons accounts only once and
all other interactions correspond to the interelectron
correlation which should be accounted necessary times. However situation
changes when the interaction of atomic electrons with projectile is
considerably large than the correlation of atomic electrons. In this case
the many-electron transition should be considered ~\cite{art2,mcg87,art3} as a result of direct
interaction of projectile's strong field. Formulae (~\ref{eq:af1},~\ref{eq:af2}) and (~\ref{eq:af3})
correspond ~\cite{art3} to the such a mechanism of direct interaction.
Below we give
the expressions for the total cross sections of one- and two-electron
transitions from ground state of nonrelativistic heliumlike atom in the
collision with relativistic highly charged ion, obtained in the large
impact parameter approximation (~\ref{eq:af4}). In all the cases
two-electron states of the heliumlike atom are described in the form of
symmetrized product of hydrogenlike one-electron wave functions. In
order to avoid orthogonolization procedure (which is not simply
defined) we choose  the one-electron hydrogenlike wave functions with
effective charge equal: $Z_{1}$ - for one-electron transitions, $Z_{2}$
-for two-electron transitions. Let's denote via $|n_1, n_2>$
two-electron states of heliumlike atom with two set of one-electron
hydrogenlike quantum numbers $n_1$ and $n_2$. Then the cross section
(~\ref{eq:af3}) of transition from the ground state $|0, 0>$ to state
$|n_1, n_2>$ in the large impact parameter approximation (~\ref{eq:af4})
is
\be
\sigma=\int d^2 b
\mid<{n_1},{n_2}\mid
exp\{-i{{\bf q}}({{\bf r}_{1}}+{{\bf r}}_{2})\}\mid0,0>\mid^{2}\,.
\label{eq:af251}
\ee
Thus the cross section is expressed by the integral (over the impact parameter)
from the product of the well known hydrogenlike formfactors ~\cite{art14}.
The cross sections of two-electron transitions (when
the states of both electron change) can be obtained directly from
(~\ref{eq:af251}) by integrating over the hole plan of impact parameters.
The cross sections of inelastic processes which include one-electron
transitions (for example, single excitation or ionization, total
inelastic cross sections) can be obtained by matching with perturbation
theory. Therefore these formulae logorithmically depend on the projectile
velocity and relativistic factor $\gamma=1/\sqrt {1-v^2/c^2}$.

The single ionization cross section when one of the electrons get into the
continuum and another one get into one of the discrete states is
\be
\sigma^{1+} =16\pi \frac{Z^2}{v^2} 0.283 \left\{\frac{1}{Z^2_1}
\left[\ln\left(\frac{5.08v^2}{Z Z_1\sqrt{1-\beta^{2}}}\right)
-\frac{\beta^2}{2}\right]-\frac{1}{Z^2_2}\ln3.72
\right\}.
\label{eq:af17}
\ee
The total cross section of the one-electron excitations of discrete spectrum
when one of the atomic electrons excites to the one of states of the one-electron
discrete spectrum, another one remains in the ground state is
\be \sigma^{1*}=16 \pi\frac{Z^2}{v^2}\frac{0.375}{Z^2_1}
\left\{\ln\left(\frac{0.256v^2}{ZZ_1 \sqrt{1-\beta^{2}}}\right)
-\frac{\beta^2}{2}\right\}.
\label{eq:af18}
\ee
The total double ionization cross section  can be obtained by summing
(~\ref{eq:af251}) over all $n_1$ and $n_2$ corresponding to the two-electron
continuum:
\be
\sigma^{2+}=16\pi\frac{Z^{2}}{v^{2}} 0.283 \frac{1}{Z^2_2} \ln{3.72}
= 9.36 \frac{Z^2}{(Z_{2})^2 v^2}.
\label{eq:af20}
\ee
The total cross section of transition of heliumlike atom to any doubly excitated
state, after summing over $n_1$ и $n_2$ (belonging to
the two-electron discret spectrum) is
\be
\sigma^{2*}=16\pi\frac{Z^{2}}{v^{2}} 0.283 \frac{1}{Z^2_2} \ln{1.15}
=2.03\frac{Z^{2}}{v^{2} Z^2_2}\,.
\label{eq:af21}
\ee
The above cross sections are connected by the general relation
\be
\sigma_r=\sigma^{1+}+\sigma^{1*}+\sigma^{2+}+\sigma^{2*}\,,
\label{eq:af22}
\ee
where the total inelastic cross section  $\sigma_{r}$, corresponds to
the arbitrary excitation of heliumlike atom and is given by
\be \sigma_{r}= 16\pi\frac{Z^2}{v^2} 0.717
\frac{1}{Z^2_a}\left\{\ln\left(\frac{1.03v^2}{Z Z_a\sqrt{1-\beta^{2}}}\right)
-\frac{\beta^2}{2}\right\},
\label{eq:af19}
\ee
where $Z_{a}$ is the effective charge of heliumlike atom in the ground
state ($1S^2$) and equal to the charge of bare nucleus minus $5/16$.

As an another example of two-electron transition to the discrete state
we give the excitation cross sections of autoionization states of
heliumlike atom with principial quantum number $n=2$ ($L$-shell). Since
in the considered collisions the spins of the electrons can not change,
the excitations of the following autoionization states are possible :
$2s^{2}\; ^{1} S$, $2s2p\; ^{1} P$, $2p^{2}\; ^{1} S$ $2p^{2}\; ^{1}
D$.

The corresponding cross sections are
$$\sigma(2p^2\; ^1D) = 2\sigma(2p^2\; ^1S) = \frac{10}{3}\sigma(2s2p\; ^1P) =
30\sigma(2s^2\; ^1S)
= \pi \frac{Z^{2}}{v^2 Z_{2}^{2}}\frac{2^{31}}{3^{19}}\frac{1}{11}.$$

In the table 1  the experimental data
(for excitation of autoionization states), the results of our
calculations and the results of calculations from ~\cite{art20} are compared.
The results are given for the sum of  $(2s2p+2p^2)$ excitaion cross sections for helium atom;
Column 1 is the
projectile energy per nucleon; column 2 is the
projectile charge; column 3 is the experimental result; column 4 is the our
result ($Z_{2}=1.97$); column 5 is the result of numerical calculations from
~\cite{art20}.

In Fig.2 the results of experiments from ~\cite{art21}, the
results of calculations obtained using
formulae (~\ref{eq:af20}) and (~\ref{eq:af17}) for the cross section of
double ($Z_{2}=1.97$) and single ($Z_{1}=1,37$) ionization of helium
atom in the collision with $U^{90+}$  (with energies 60, 120 и 420 Mev/n)
as well as the ratio $\sigma^{2+}/\sigma^{1+}$ are given. Correctness of
the choice for values of the effective charges $Z_{1}=1.37$ and $Z_{2}=1.97$
is confirmed by the good accordance of our results given in table 2 with
the experimental ones from ~\cite{knu84,mcg87}.

The above data for the
single ionization cross section ($\sigma^{1+}$) as well as the sum of cross sections $\sigma^{1+}$ and $\sigma^{2*}$  
can be used for the estimations of total
cross section of formation of one-electron helium ions as a result of
direct ionization and Auger decay of various doubly excited states of
helium atom, since for light atoms Auger decay is the dominating decay
channel of doubly excited states (excluding  comparitively  small
number of doubly excited states for which Auger decay is forbidden
by selection rules ~\cite{wen27,pro60,bet60}).

\section*{Excitation and ionization of nonrelativistic complex atoms}

Though strong field of highly charged ion leads to high ionization
probabilities, in the case of multiple ionization and excitation
the large impact parameter approximation (~\ref{eq:af4}) can break
down, due to the fact that corresponding cross section may become
comparable with atomic sizes. Therefore
more general consideration on the ground of formula
(~\ref{eq:af3}) is needed. Let's consider the electrons of nonrelativistic
(before and after collision) multielectron atom as
distinguishable and each electron described by hydrogenlike wave
function. Then the initial wave function. $\Psi_{0}({{\bf
r}}_{1},\ldots,{{\bf r}}_{N_{0}})=
\prod^{N_{0}}_{i=1}\phi_{i}({{\bf r}}_{i})\:,\;\;$
final one is $\Psi_{f}({{\bf r}}_{1},\ldots,{{\bf r}}_{N_{0}})=
\prod^{N_{0}}_{i=1}\psi_{i}({{\bf r}}_{i})\:.$
Therefore for the total ($N_{0}-N$) - fold ionization probability of
nonrelativistic $N_{0}$ - electron atom, corresponding to the
ionization of ($N_{0}-N$) electrons by simultaneous
transition of other $N$ electrons to any of states of discrete
spectrum, with account of unitarity, according to (~\ref{eq:af3})
we have
\be
\;\;\;\;\;\;W^{(N_{0}-N)+}({\bf b})=\frac{N_{0}!}{(N_{0}-N)!N!}\prod^{N_{0}-N}_{i=1}
p_{i}({\bf b})
\prod_{j=N_{0}-N+1}^{N_{0}}(1-p_{j}({\bf b}))\;,
\label{eq:af23}
\ee
where $\prod_{j=N_{0}-N+1}^{N_{0}}(\ldots)=1\;,\;\; for \;\;N=0$;
\be
p_{i}({\bf b})=\int d^{3}k_{i}\mid d^{3}r_{i}
\psi^{*}_{{{\bf k}}_{i}}({{\bf r}}_{i})
exp\{-i\chi_{i}({\bf b},{\bf r}_{i})\}\phi_{i}({{\bf r}}_{i})\mid^{2},
\label{eq:af24}
\ee
${\bf k}_{i}$ - is the momentum of
${i}$ - th electron in the continuum.
This probability depends on the vector ${\bf b}$ but after
averaging over the orbital moment of the initial state of atom it
will depend on only $\mid {\bf b}\mid$. Let's introduce avareged
over moment $l$ and its projection  $m$ inelastic formfactor for
each electron, which also averaged over all atomic shells:
\be
p(b)=\frac{1}{n_{0}}\sum_{n=1}^{n_{0}}\frac{1}{M_{n}}\sum_{l,m}
\int d^{3}k\mid\int d^{3}r\psi_{{\bf k}}^{*}
({\bf r})exp\{-i\chi({\bf b},{\bf r})\}\phi_{nlm}({\bf r})\mid^{2}\;,
\label{eq:af25}
\ee
where summation carryied out over the all values of $l$ and $m$ of
$n$ - th shell, $M_{n}$ - is the number of such values,
$n$ - is the principial quantum number,
$n_{0}$ - is the number of atomic shells. It is obvious that
$p(b)=p(\mid {\bf b}\mid)$ not depends on the angles of vector
${\bf b}$, i.e. $p(b)$ has the sense of average one-electron ionization
probability. Then replacing in (~\ref{eq:af23}) each one-electron
formfactor with average value from (~\ref{eq:af25}) we get, for
the ionization probability of ($N_{0}-N$) electrons, the usual expression of
independent electron approximation ~\cite{art2,art8}. However the
effective charge $Z^{*}$ of the atomic nucleus depends on the
ionization degree. In order to account this fact we make in
(~\ref{eq:af25}) the following substitutions
${\bf k}={\bf k}/Z^{*}$,
${\bf b}={\bf b}Z^{*}$, ${\bf r}={\bf r}Z^{*}$,
corresponding to the transition to Coulomb units ~\cite{art14}.
Then left side of (~\ref{eq:af25}) can be calculated using the
wave functions of atomic hydrogen with unit charge. Below
$p(b)$ will mean the avareged by (~\ref{eq:af25}) hydrogen atomic
fromfactor. Such a replacement enables to calculate the ionization
cross section for general (than independent electron
approximation) cases. Consider ionization of high
multiplicities $N_{0}\gg1$, $N_{0}-N\gg 1$.
For the ionization of
$N_{0}$ electrons, ($N=0$ in (~\ref{eq:af23}))  $W$ can be
reduced to the product of $N_{0}$ one-electron formfactors. We
introduce the effective charge $Z^{*}_{N_{0}}$ of the nucleus
corresponding to the total ionization of the atom. Replacing each
one-electron formfactor with the average
(~\ref{eq:af25}) one obtains the total ionization probability
$W^{N_{0}+}=\left[p(b)\right]^{N_{0}}\;,$
where $b=bZ^{*}_{N_{0}}$. Integration in (~\ref{eq:af3})
over $d^2 b$ can be performed asymptotically
($N_{0}\gg1$) by the Laplace method assuming that
$p(b)$ has only one maximum for the $b=b_{1}=0$. Existence of this
maximum follows from ~\cite{art13,art7}. As a result the total
$N_{0}$ - fold ionization cross section is
\be
\sigma^{N_{0}+}=\pi\frac{1}{(Z^{*}_{N_{0}})^{2}}
\left[\frac{-2\pi}{p''(b_{1})N_{0}}\right]^{1/2} \left[p(b_{1})
\right]^{N_{0}+1/2}\;,
\label{eq:af26}
\ee
here and below $b_{1}$ is the maximum point of the function
$p(b)$, $p''(b_{1})$ is the second derivative of
$p(b)$ over ${b^2}$.

In the case $(N_{0}-1)$ - fold ionization the ionization probability
is the difference of two terms one of which contains the product
$N_{0}-1$ one-electron formfactors (corresponding effective
charge is $Z_{N_{0}-1}^{*}$); second term is the product of
$N_{0}$ one-electron formfactors and corresponding effective
charge is $Z^{*}_{N_{0}}$. Integrating each term by the Laplace method
 we obtain the cross section of $(N_{0}-1)$ - fold ionization.
\be
\sigma^{(N_{0}-1)+}=N_{0}\sigma^{N_{0}+}
\left[\left(\frac{Z^{*}_{N_0}}{Z^{*}_{N_{0}-1}}\right)^{2}
\left(\frac{N_{0}}{N_{0}-1}\right)^{1/2}\frac{1}{p(b_{1})}-1\right]\;.
\label{eq:af27}
\ee
Analogously in the general case of $(N_{0}-N)$ - fold ionization acting
analogous one obtains
\begin{eqnarray}
\sigma^{(N_{0}-N)+}=\frac{N_{0}!\sigma^{N_{0}+}}{(N_{0}-N)!\,N!}
\sum_{m=0}^{N}(-1)^{m}\left(\frac{Z^{*}_{N_{0}}}
{Z^{*}_{N_{0}-N+m}}\right)^{2}\times \nonumber\\
\;\;\;\;\times\frac{N!\sqrt{N_{0}/(N_{0}-N+m)}}{(N-m)!\,m!}
\{p(b_{1})\}^{-N+m}\;,
\label{eq:af28}
\end{eqnarray}
where $Z^{*}_{N_{0}-N+m}$ is the effective charge for
$(N_{0}-N+m)$ - fold ionization.

Obtained formulae (~\ref{eq:af26}), (~\ref{eq:af27}) and (~\ref{eq:af28})
enables one to calculate the ionization cross section of any
multiplicity (when $N_{0}\gg1,\;(N_{0}-N)\gg1$), or to construct
other cross section using known two experimental values of cross
section. It is simplest to find $\sigma^{N_{0}+}$ and
$\sigma^{(N_{0}-1)+}$ using which one can obtain
$p(b_{1})$ and substitute it into (~\ref{eq:af28}). Thus
the cross section of ($N_{0}-N$) - fold ionization is expressed
via $\sigma^{N_{0}+}$ and $\sigma^{(N_{0}-1)+}$. The results of such
calculations for multiply (up to 8) ionization of $Ne$ and $Ar$ (up
to 18) atoms are given in Figs. 3 - 5. In this calculations the
effective charge taken equal to the ionization degree ($Z^{*}_{N}=N$).
As is seen from these Figs. accordance of our
calculations with the experimental data from
~\cite{art22,art23} is good enough even for the ionization of
lower multiplicities, lying out of the region of applicablity
 ($N_{0}-N\gg1$) of formulae (~\ref{eq:af26}), (~\ref{eq:af27})
and (~\ref{eq:af28}).

\section*{K-vacancy production in the collisions with heavy atoms}

In this section the above results on the eikonal approximation are applied
for the calculation of probabability and cross section of K-vacancy production
in the collision of relativistic highly charged ions with heavy (relativistic)
atoms. As was mentioned above the integrand in (\ref{eq:af3}) can be interpreted
as the probability of transition
from state $\mid\psi_{i}>$ to state $\mid\psi_{f}>$ in the collision with the
impact parameter ${\bf b}$.
Direct calculation of the probability can be performed only numerically.
Therefore we assume the following simplifications:\\
a) We will use the large impact parameter approximation (\ref{eq:af4})\\
b) The target atom is considered as a quasirelativistic i.e.,\\
 $exp[i\sum_{a} x_{a}v(E_{f}-E_{i})/c^2] \approx1$;
 $\gamma^{-N}S^{-2} \approx1,$\\ that means
as a wave functions   $\mid\psi_{i}>$
and $\mid\psi_{f}>$ one can use the well known Darwin and Zommerfeld-Mau
wave functions ~\cite{art4}.\\
Then for the K-vacancy production probability
we have
$$ I(b) = N\int a_{k}^2 dk,$$
where
$$ N = (1+\frac{Z_{a}^2\alpha^2}{4})^{-1}, (\alpha=1/137) $$

$$a_{n}^2 =
N\frac{2^{8}Z_{a}^6kq^2(q^2+(Z_{a}^2+k^2)/3)exp[-\frac{2Z_{a}}{k}]
arctg[\frac{2Z_{a}k}{q^2+Z_{a}^2-k^2}]dk}{(1-exp(-\frac{2\pi Z_{a}}{k}))
((q+k)^2+Z_{a}^2)^3
((q-k)^2+Z_{a}^2)^3(1+k^2\alpha^2)}$$
is the square of the absolute value of
 the well known relativistic hydrogenlike formfactor $<{\bf k}|exp\{i{\bf q}{\bf r}\}|i>\;$
 integrated over the  emission angles of the electron
~\cite{a4,art4}.
It differs from the
nonrelativistic one by the presence of the constant N and the factor
$(1+k^2\alpha^2)^{-1}$, ($Z_{a}$ is the charge of target).
The ionization probability (as a function of impact parameter)
calculated using this formula for  $U^{92+}$-$U^{91+}$ collision
is given in Fig.6. As is seen from this figure our approach based on
the eikonal approximation gives, for small impact parameters, an ionization
probability  which is less than unity.
Let's calculate now the K-vacancy production cross section
using the formula (~\ref{eq:afsig}).
The above simplifications a) and b) leads to  the formula (~\ref{alpha})
which is of the same form as the formula (~\ref{alphan}) but
in this formula as the wave  functions $\mid\psi_{i}>$
and $\mid\psi_{f}>$ are taken Darwin and Zommerfeld-Mau
wave functions. Therefore in the calculations of
 $\lambda_{i}$, $\alpha_{i}$ and $\omega_{i}$ the dependence on  $Z_a$
cannot be factorized and the quantities
 $\lambda_{i}$, $\alpha_{i}$ and $\omega_{i}$ as a functions of  $Z_a$ should be
calculated nimerically. As an example of such calculation
in Fig. 7 the dependence of K-vacancy production
cross section on the relativistic factor $\gamma$ is given for
$U^{92+}$-$U^{91+}$ collision .

\section*{Conclusion}
Thus we have derived general formulas for cross section which are applicable
in the case of collisions of atoms with ions of arbitrary charge.
These obtained formulae are applied to the (analitically and numerical) calculations
of the following processes:
1) the excitation and ionization
cross sections of (nonrelativistic) hydrogenlike and heliumlike atoms in the
collsions  with relativistic highly charged ions;\\
2) the probability and cross section of K-vacancy
production in the relativistic $U^{92+}$-$U^{91+}$ collision;\\
3) the multiple (up to 8 for Ne and up to 18 for  Ar) ionization cross sections
in the collision of complex atoms with relativistic highly
charged ions.

We also have obtained simple analytical expressions for inelastic
cross sections and derived recurrence relations between the cross sections
of various multiplicities. Obtained theoretical results are compared
with the experimental data.

Besides that the above calculations of cross section and ionization probability,
using the eikonal approximation, enables to
avoid some difficulties appearing in the case of application of
perturbation theory to the relativistic highly charged ion -heavy atom
collisions and leads to the result coinciding, in the ultrarelativistic limit,
with the known exact one.

\newpage

\begin{tabular}{|c|c|c|c|c|}                                             \hline
Energy,    & Charge  & Experiment,    & Our results,   &
Calculation ~\cite{art20},    \\
MeV/n.     &  ion   &  $10^{-19}cm^2$ &$10^{-19}cm^2$ & $10^{-19}cm^2$ \\ \hline
1.84       &   6    &$8.305 \pm 1.744$&  18.45        &  25.6          \\
1.5        &   6    &$20.1 \pm 7.20$  &  22.61        &  31.8          \\
1.5        &   9    &$48.99 \pm 17.66$&  50.79        & 111.6          \\
\hline
\end{tabular}
\newline

{\bf Table 1.} The sum of the excitation cross sections of the autoionization states $2s2p\ ^{1}P$ and
$2p^{2}\ ^{1}D$ ($\times10^{19}\;\; см^{2}$) for the helium atom.

\vskip 2cm

\hskip 2cm

\begin{tabular}{|c|c|c|c||c|c|c|} \hline
  Energy,  &   Charge &  $\sigma^{2+}\,$ & $\sigma^{2+}\,$ &  $\sigma^{1+}\,$ & $\sigma^{1+}\,$& $\sigma^{1+}+\sigma^{2*}\,$  \\
  MeV/n.   &   ion    &  experiment  &    theory          & experiment
&  theory   & theory  \\ \hline
   0.64    &   8      &  1.32         &  1.687             &  7.9          &  10.231  &  10.597   \\  \hline
   1.00    &   8      &  1.06         &  1.08              &  6.7          &  8.11  &  8.344  \\ \hline
   1.44    &   8      &  0.45         &  0.75              &  5.9          &  6.518 &  6.68   \\ \hline
    1.4    &   15     &  2.91         &  2.712             &  17.9         &  17.798  &  18.385    \\ \hline
    1.4    &   18     &  4.50         &  3.905             &  22.4         &  23.322  &  24.168  \\ \hline
    1.4    &   20     &  5.41         &  4.821             &  26.0         &  27.146  &  28.191  \\ \hline
    1.4    &   36     &  16.0         &  15.621            &  57.2         &  58.206  &  61.59  \\ \hline
    1.4    &   37     &  16.8         &  16.501            &  59.5         &  60.02   &  63.594  \\ \hline
    1.4    &   44     &  23.0         &  23.335            &  72.1         &  71.779  &  76.833  \\ \hline
\end{tabular}
\newline

{\bf Table 2.} The cross sections of double and single ionization of the  helium atom.

\newpage
\large
\centerline{\bf Figure captions}
\begin{itemize}

\item[Fig.1] The ionization cross section of hydrogen atom by
$C^{6+}$ ions obtained using: \\
1 - The Born approximation, \\
2 - The method of matching (formulae (~\ref{eq:af10})), \\
3 - The Glauber approximation ~\cite{art9}, \\
4 - The sudden approximation ~\cite{art7}.

\item[Fig.2] Experimental results from ~\cite{art21} and results
of calculations (by formulae (~\ref{eq:af20}) and (~\ref{eq:af17}))
for single and double ionization cross section of heliumlike atom
by  $U^{90+}$ ions with energy
60, 120 и 420 Mev/n., as well as for the ratio
$\sigma^{2+}/\sigma^{1+}$: \\
$\bullet$ - experiment, $  {\times}$ - calculations.

\item[Fig.3] The dependence of the multiply ionization cross section for $Ar$ atom
colliding with  $U^{75+}$ ions by energy 15 Mev/n:
\newline  {$\Box$ - experiment from ~\cite{art22},}
\newline {$\triangle$ - our results.}

\item[Fig.4] The experimental results from ~\cite{art22,art23} and
calculations (using (~\ref{eq:af26}) - (~\ref{eq:af28})) for the
multiply ionization cross sections of $Ne$ atoms in the collision
with 120 Mev/n. $U^{90+}$ ions as a function of ionization degree
$n$:
\newline  { $\Box$ - experiment,}
\newline  {$\triangle$ - our results.}

\item[Fig.5] Experimental results from ~\cite{art22,art23} and
results of calculations (using (~\ref{eq:af26}) - (~\ref{eq:af28}))
for multiply ionization cross sections of $Ar$ atoms in the
collisions with 120 Mev/n. relativistic  $U^{90+}$ ions as a
function of ionization degree $n$:
\newline   {$\Box$ - experiment,}
\newline  {$\triangle$ - our results.}

\item[Fig.6] K-vacancy production probablity as a function of the impact
parameter $b$  for $U^{92+}$ с $U^{91+}$ collision. Solid line -our result,
dashed line is the result of Valluri S.R. et.al ~\cite{vall84}.
.

\item[Fig.7] K-vacancy production cross section as a function of relativistic factor
 $\gamma$ for  $U^{92}+U^{91}$ collision.
Solid line is the result of our calculations, dashed one is the result from
~\cite{art4}. The cross section is given in barns.

\end{itemize}

\end{document}